\documentclass[prd,aps,showpacs,showkeys]{revtex4-1}

\usepackage{bm}
\newcommand{\bb}{\begin{eqnarray}}
\newcommand{\ee}{\end{eqnarray}}

\begin{document}

\title{ \bf  Free and bound spin-polarized fermions in the fields of
Aharonov--Bohm kind}
\author{V.R. Khalilov, I.V. Mamsurov and Lee Ki Eun}
\affiliation{Faculty of Physics, Moscow State University, 119991,
Moscow, Russia}

\begin{abstract}
The scattering of electrons by an Aharonov--Bohm field is
considered from the viewpoint of quantum-mechanical problem of
constructing a self-adjoint Hamiltonian for the Pauli equation.
The correct domain for the self-adjoint Hamiltonian, which takes
into account explicitly the electron spin is found. A
one-parameter self-adjoint extension of the Hamiltonian for
spin-polarized electrons in the Aharonov--Bohm field is selected.
The correct domain of the self-adjoint
Hamiltonian  can contain  regular and
singular (at the point ${\bf r}=0$) square-integrable functions
on the half-line with measure $rdr$. We argue that the physical
reason of the existence of  singular functions is the additional
attractive potential, which appear due to the interaction
between the  spin magnetic moment of
fermion and Aharonov--Bohm magnetic field.  The scattering amplitude and cross section
are obtained for spin-polarized electrons scattered by the
Aharonov--Bohm field. It is shown that in some range of
the extension parameter  there appears a bound state.
Since the Hamiltonian of
the nonrelativistic Dirac--Pauli equation for a massive neutral
fermion with the anomalous magnetic moment (AMM) in the electric
field of a linear charge aligned perpendicularly to the fermion
motion  has the form of the Hamiltonian for the Pauli equation in
the Aharonov--Bohm flux tube, we also calculate the scattering
amplitude and cross section for the neutral fermion.
\end{abstract}



\maketitle

\section{Introduction}

The quantum Aharonov--Bohm (AB) effect, predicted in
\cite{1}, is an important physical phenomenon.  Due to the
cylindrical symmetry of Aharonov--Bohm field configuration, the
relevant quantum mechanical system is invariant along the symmetry
($z$) axis and essentially two-dimensional in the $xy$ plane. So,
the models applied to describe  AB effect can usually be reduced
to the (2+1)-dimensional ones.  Such models are studied by means
of the Dirac equation in 2+1 dimensions and results obtained are
applied to other problems. Solutions to the two-component Dirac
equation in the AB potential were first obtained and discussed by
Alford and Wilczek in Ref. \cite{aw} in a study  of the
interaction of cosmic strings with matter and the production of
particle-antiparticle pairs by the vector potential of a moving
cosmic string. The effect of vacuum polarization in the field of
infinitesimally thin solenoid  is recently investigated in
\cite{jmpt} and a wonderful phenomenon is revealed: the induced
current is finite periodical function of the magnetic flux.

 The scattering of spinless charged particle by an
infinitely long thin solenoid was examined by Tyutin in \cite{ivt}
from the viewpoint of quantum-mechanical problem of constructing
of self-adjoint Hamiltonians. The Hamiltonian for the
Aharonov--Bohm problem is essentially singular and hence it cannot
be immediately defined in the domain $[0,\infty)$ for any
differentiable and enough rapidly decreasing (at $r\to \infty$) functions
in the Hilbert space of square-integrable functions
on the half-line with measure $rdr$ \cite{ivt}. Usually, the Hamiltonians are
symmetric operators in its natural domain. The problem of
constructing a self-adjoint Hamiltonian is to find all
self-adjoint extensions of given symmetric operator and then to
select correct self-adjoint extension. The correctness of the
known Aharonov--Bohm solutions for scattering problem was analyzed
for spinless particles in \cite{ivt}, in which a self-adjoint extension of the
Hamiltonian is selected by physical condition -``the principle of
minimal singularity'': the Hilbert-space functions for which the
Hamiltonian is defined must not be singular.

A one-parameter self-adjoint extension of the Dirac Hamiltonian in
2+1 dimensions in the pure AB field  was constructed by means of
acceptable boundary conditions in \cite{phg,ampw,phgj,khnew}. For some
extension parameters a domain of the self-adjoint Hamiltonian was
found to exist in which the Hamiltonian is self-adjoint and
commutes with the helicity operator \cite{cfp}. One- and
two-parameter families of self-adjoint Dirac Hamiltonians in the
superposition of the Aharonov--Bohm solenoid field and a collinear
uniform magnetic field in the respective  2+1 and 3+1 are
constructed in \cite{ggs,ggsv}.

 Nevertheless, there remains a question about the correctness of solutions
for the scattering of spin particles in the AB flux tube. Indeed,
for this problem we need to use the Hamiltonian for the Pauli
equation, which contains one more extremely singular (spin) term
taking account of the interaction between the spin magnetic moment
of electron and magnetic field.

An interesting and important corollary to the Aharonov-Bohm
geometric phase is a phase acquired by the wave function of a
neutral massive fermion with the magnetic moment when it
propagates in an electric field of a uniformly  charged long
conducting thread aligned perpendicularly to fermion motion (such
a field will be called the AC configuration). The fermion
transport is influenced by the phase acquired by the fermion wave
function and the resulting phase difference leads to a spin-field
dependent effects in scattering (the Aharonov--Casher (AC) effect
\cite{ahc}). The Hamiltonian of nonrelativistic Dirac--Pauli
equation for a massive neutral fermion with the anomalous magnetic
moment in the electric field of a linear charge aligned
perpendicularly to the plane of fermion motion has the same form
with the Hamiltonian for the Pauli equation in the pure AB
potential. The nonrelativistic quantum motion of an uncharged
massive fermion with an anomalous moment interacting with the
electric and magnetic fields produced by linear electric and
magnetic sources in a conical spacetime was studied in
\cite{erbm}.

Thus, the self-adjoint extensions for the nonrelativistic
Hamiltonians with the particle spin (this term leads to additional
singular $\delta({\bf r})$ potential in the Hamiltonian) have
still not been constructed.

This paper is organized as follows.  In Section II we find a
one-parameter self-adjoint extension of the Hamiltonian for
spin-polarized electrons in the Aharonov--Bohm field.
In Section III we find the scattering amplitude and cross
section  as well as the wave function of bound state
for spin-polarized electrons in the Aharonov--Bohm flux
tube of zero radius. In Section IV we discuss
the scattering problem for spin-polarized neutral massive
fermion with AMM in the field of uniformly charged conducting
thread of zero radius. In Section V we briefly discuss physical
results.

We shall adopt the units where $c=\hbar=1$.

\section{Self-adjoint extensions for the Pauli Hamiltonian in an
AB potential}

The Pauli equation for an electron of mass $m$ and charge $e<0$ in
an AB potential specified in cylindrical coordinates as
 \bb
  A^0=0,\quad A_r=0,\quad A_{\varphi}=\frac{B}{r}, \quad A_z=0,
\quad  B=\frac{\Phi}{2\pi}, \label{eight1}\\
\quad r=\sqrt{x^2+y^2}, \quad
\varphi=\arctan(y/x)\phantom{mmmmmmmmm} \nonumber
 \ee
 is
 \bb
i\frac{\partial}{\partial t}\Psi(t,{\bf r},z)={\cal H}\Psi(t,{\bf
r},z),\quad {\bf r}=(x, y), \label{eq12} \ee where $\Psi(t,{\bf
r},z)$ is a spinor and the Hamiltonian ${\cal H}$ is  \bb 2m{\cal
H}= -\frac{\partial^2}{\partial
r^2}-\frac{1}{r}\frac{\partial}{\partial r} -
\frac{1}{r^2}\left(\frac{\partial^2}{\partial \phi^2} +
2i|e|B\frac{\partial}{\partial \phi} -
e^2B^2\right)-\frac{\partial^2}{\partial z^2} +|e|B\sigma_3
\pi\delta({\bf r}) \label{e12}
 \ee

Here $\sigma_3$ is the Pauli spin matrix
$$
\sigma_3=\left(\begin{array}{cc}
1 & 0\\
0 &-1\\
\end{array}\right).
$$
The last term in (\ref{e12}) describes the interaction of the
electron spin with a magnetic field.

Equation (\ref{eight1}) is the potential of an infinitely thin
solenoid with a finite magnetic flux $\Phi$ in the $z$ direction
in the range $r>0$. The magnetic field ${\bf H}$ is restricted to
a flux tube of zero radius \bb
 {\bf H}=(0,\,0,\,H)=\nabla\times {\bf A}= B\pi\delta({\bf r}).
\label{e1s} \ee Therefore, we need to construct a self-adjoint
Hamiltonian of Eq. (\ref{e12}) in cylindrical coordinates for the
electron motion in the $xy$ plane. It is seen that the inclusion
of particle spin in the AB problem leads to the additional
$\delta({\bf r})$ potential in the Hamiltonian \cite{khal}.

We seek solutions to the Pauli equation for the electron motion in
the $xy$ plane in the form
 \bb
 \Psi(t,
r, \varphi) &=&\exp(-iEt) f(r, \varphi)\psi \equiv
 \exp(-iEt)\sum\limits_{l=-\infty}^{\infty}F_l(r)\exp(il\varphi)
\psi^s, \label{three}
 \ee
where $E$ is the electron energy, $l$ is an integer, and $\psi^s$
is a constant two-spinor
 \bb \psi^s = \left(
\begin{array}{c}
\psi^1\\
\psi^{-1}
\end{array}\right) = \frac{1}{2}
\left( \begin{array}{c}
1+s\\
1-s
\end{array}\right),
\label{1four}
 \ee
where $s=\pm 1$ characterizes the electron spin projection on the
$z$ axis.

The radial part $F_l(r)$ of the wave function satisfies
 \bb
 \left(\frac{d^2}{dr^2}+\frac{1}{r}\frac{d}{dr} - \frac{l^2}{r^2} -
\frac{2|e|Bl}{r^2} - \frac{e^2B^2}{r^2} +
2Em-|e|Bs\frac{\delta(r)}{r}\right)F_l(r)=0 \label{e13}
 \ee
and it depends explicitly on the number $s$, which selects a
particular value of the spin projection on the $z$ axis.

At first, it is rewarding to remind the problem of constructing
solutions for the linear differential Bessel expression for
functions $F_l(r)$ or $f(r)=F_l(r)/\sqrt{r}$ of the forms
 \bb
 l(F)= \left(-\frac{d^2}{dr^2}-\frac{1}{r}\frac{d}{dr} + \frac{(l+\mu)^2}{r^2}\right)F_l(r), \label{3e23}
 \ee
or
 \bb
 l(f)= \left(-\frac{d^2}{dr^2} + \frac{(l+\mu)^2-1/4}{r^2}\right)f(r), \label{e23}
 \ee
where $\mu=|e|B$. In the form (\ref{e23}) the quasiderivative
$l(f)$ is self-adjoint \cite{nai,crein} and can be defined in the
range $-\infty <x<\infty, \quad x\equiv r$.

In the Hilbert space $L_2(a, b)$, i.e. the space square integrable
functions in the range $(a,b)$ (the interval $(a,b)$ may be
infinite)  the linear everywhere dense set $D_0^{\prime}$ is
considered. The set $D_0^{\prime}$ consists of the infinitely
differential functions that vanish outside some cut $[c,d]$ (the
cut may be various for various functions) contained completely in
the range $(a, b)$. On $D_0^{\prime}$ the operator $L_0$ is
defined by equality $L_0f=l(f)$ \cite{nai,crein}. In each cut
$[c,d]$ the Lagrange identity
 \bb
 \int\limits_{c}^{d}l(f)\bar g dr - \int\limits_{c}^{d}f l(\bar g)dr =
 \left(f\frac{d\bar g}{dr} - \frac{df}{dr} \bar g\right)_c^d\equiv [f,g]^d_c=0 \label{e33}
 \ee
is valid for any function $\bar g$ being complex conjugate of $g$.
So $L_0$ is a symmetric operator. The closure $A_0$ of operator
$L_0$ will be also a symmetric operator \cite{nai,crein}.
 We need to construct self-adjoint extensions  $A^{\theta}$
of operator $A_0$.
The adjoint operator $A_0^*$ is
given by equation \bb A_0^*f=l(f).\label{adj}\ee  $A_0^*$ is
defined on the set $D(A_0^*)$ of all functions that are absolutely
continuous with their first derivative in the range $(a,b)$ and
the quasiderivative $l(f)$ belonging to $L_2(a, b)$.

If Eqs. (\ref{3e23}), (\ref{e23}) are considered in the range
$(0,\infty)$ then they will be singular  on both ends. In this
case the domain $D(A_0)$ of operator $A_0$ contains all functions
$f$ from $D(A_0^*)$ for which \bb [f,g]^b_a=0 \label{3e33} \ee for
all $g$ from $D(A_0^*)$. In the most important case   $\mu$ is
nonintegral and the corresponding operator $A_0$ will have the
deficiency indices (2,2) at $0<|l+\mu|<1$. The eigenfunctions
$G^{\pm}(r)=g^{\pm}\sqrt{r}$ of the adjoint operator $A_0^*$ with
the eigenvalues $\pm i$ satisfy equations
 \bb
 \left(\frac{d^2}{dr^2}+\frac{1}{r}\frac{d}{dr} - \frac{(l+\mu)^2}{r^2} \mp
ip^2\right)G_l^{\pm}(r)=0, \label{eadj}
 \ee
where $p^2\ne 0$ is inserted for dimensional reasons. Square
integrable normalized solutions of (\ref{eadj}) are expressed by
the Bessel functions of third kind (see, for instance
\cite{GR,olv})\bb G_l^+(r)= A^+_lH_{\nu}^{(1)}(e^{
i\pi/4}pr),\quad G_l^-(r)= A^-_lH_{\nu}^{(2)}(e^{-
i\pi/4}pr),\label{func2} \ee where $A^{\pm}_l$ are the
normalization factors  and \bb
 \nu =|l+\mu|.
\label{not12}
 \ee

 If $\mu$ is nonintegral then the
operator $A_0^*$ has two eigenfunctions belonging to
$L_2(0,\infty)$ for each eigenvalue $\pm i$. Indeed, writing $\mu$
as $\mu=[\mu]+\gamma$, where $[\mu]$ is the largest integer
$\le\mu$, it is seen that these functions belong to the critical
subspace $0<\gamma<1$ (with $l=-[\mu]-1$). The self-adjoint
extensions  of $A_0$ can be parameterized by the parameter
$\theta$. The correct domain $D(A^{\theta})$ for the self-adjoint
extension $A^{\theta}$ of $A_0$ is then given by \bb
D(A^{\theta})= D(A_0) + C_1H_{1-\gamma}^{(1)}(e^{ i\pi/4}pr)+
C_2H_{\gamma}^{(1)}(e^{i\pi/4}pr)
+B_1H_{1-\gamma}^{(2)}(e^{-i\pi/4}pr) + B_2H_{\gamma}^{(2)}(e^{-
i\pi/4}pr), \label{exten}\ee where $B_k=C_k\exp(-i\theta)$, $C_k$
are the arbitrary complex numbers and $\theta$ is an arbitrary
(but fixed for a given extension) parameter. The domain $D(A_0)$
contains absolutely continuous, square integrable on the half-line
with measure $rdr$ and regular (at the point $r=0$) functions. If
$\mu$ is a non-integer then the operator $A_0^*$ has two
eigenfunctions belonging to $L_2(0,\infty)$ for each eigenvalue
$\pm i$. Let us write (see, \cite{olv}) \bb H_{\nu}^{(1)}(z)=
\frac{i[e^{-i\nu\pi} J_{\nu}(z)- J_{-\nu}(z)]}{\sin\nu\pi},\qquad
H_{\nu}^{(2)}(z)= -\frac{i[e^{i\nu\pi} J_{\nu}(z)-
J_{-\nu}(z)]}{\sin\nu\pi}, \label{olvfun2} \ee where $J_{\nu}(z)$
and $J_{-\nu}(z)$ denote the regular and irregular Bessel
functions.  The asymptotic behavior of the Bessel function at
$r\to 0$ is \bb J_{\nu}(x)\cong
\frac{x^{\nu}}{2^{\nu}\Gamma(1+\nu)},\label{besasy}\ee where
$\Gamma(z)$ is the gamma function of argument $z$.

 Applying
(\ref{besasy}) we derive \bb CH_{\nu}^{(1)}(e^{ i\pi/4}pr)+
Ce^{-i\theta}H_{\nu}^{(2)}(e^{-i\pi/4}pr)=\frac{2C
e^{-i\theta/2}}{\sin\pi\nu}\times \nonumber\\
\times\left[\frac{2^{\nu}(pr)^{-\nu}}{\Gamma(1-\nu)}\sin\left(\frac{\theta}{2}-\frac{\pi\nu}{4}\right)-
\frac{2^{-\nu}(pr)^{\nu}}{\Gamma(1+\nu)}\sin\left(\frac{\theta}{2}-\frac{3\pi\nu}{4}\right)\right],
\label{ext1}\ee in which $\nu=\gamma, 1-\gamma$.

For any function $f(r)$ from the defect subspaces the Lagrange
identity (\ref{e33}) has to be satisfied \bb
 [f,g]^{\infty}_0=0. \label{e33defect}
 \ee
Since the functions from defect subspaces must be
square integrable  on the half-line $[0,\infty)$ with measure
$rdr$ (i.e. they, in fact, obey the Lagrange
identity at $r\to\infty$) it is enough to put a boundary condition at
the origin \bb
 \lim_{r\to 0}r\left(f\frac{d\bar g}{dr} - \frac{df}{dr} \bar g\right)=0. \label{boune33}
 \ee
 A symmetric operator
is self-adjoint if its domain coincides with that of its adjoint
(see, for instance \cite{phg}). So, one has to posit the same
boundary condition at the origin for functions from the spaces
$G^{\pm}(r)$ defined on the half-line.

The correct domain for the self-adjoint extension of the
Hamiltonian of spinless particles was selected in \cite{ivt} by
the physical condition of ``minimal singularity'' - the domain of
the Hamiltonian should be composed by entirely regular functions.
It follows from (\ref{ext1}) that this requirement fixes the
parameter  $\theta=\pi\gamma/2$ for $\nu=\gamma$ and
$\theta=\pi(1-\gamma)/2$ for $\nu=1-\gamma$. Since $0<\gamma<1$
the correct domain of the Hamiltonian is composed by entirely
regular functions for $0<\theta<\pi/2$. Insisting on regularity of
all functions at the origin forces one to reject irregular (but
square integrable on the measure $rdr$) solutions in the defect
subspaces, which can entail a loss of completeness in the angular
basis. Indeed, the inclusion of particle spin in the AB problem
leads to the presence of spin-dependent potential term ($\sim
(|e|sB/r)\delta(r)$) in the Hamiltonian, which can be taken into
account also by means of the boundary condition at the origin. Let
 $B$ be chosen, for definiteness, positive. Then this potential is repulsive for $s=1$, and
attractive for $s=-1$ what means that if $s=1$ then functions from
the defect subspaces defined on the real half-line should be
regular at the point $r=0$ but if $s=-1$ then singular
(concentrated at the origin) square integrable functions have to
appear. It should be emphasized that the appearance of
concentrated solution at the point $r=0$ is due to the physical
reason, namely, by the presence of attractive potential
$(|e|sB/r)\delta(r)$. We see from (\ref{ext1}) that the defect
subspaces does not contain the regular functions with indices
$\gamma$ and $1-\gamma$, respectively,  at $\theta=3\pi\gamma/2$
for $\nu=\gamma$ and $\theta=3\pi(1-\gamma)/2  $ for
$\nu=1-\gamma$. The range of the extension parameter in this case
is $\pi/2<\theta<3\pi/2$.

It should be noted that the linearly-independent solutions of equation \bb
 \left(-\frac{d^2}{dr^2} + \frac{(l+\mu)^2-1/4}{r^2}\right)f(r)=\lambda f \label{1e23}
\ee with the Bessel operator (\ref{e23}) are expressed as
 \bb
 f(r)=
a_l\sqrt{r}J_{\nu}(\sqrt{\lambda}r)+b_l\sqrt{r}J_{-\nu}(\sqrt{\lambda}r),
 \label{regs12}
 \ee
where $a_l$ and $b_l$ are the normalization factors and $\lambda$
 a is a real positive number. The
continuous part of spectrum of any self-adjoint extension of
operator $A_0$ (\ref{1e23}) considered in $[0,\infty)$ coincides
with the positive semiaxis  $\lambda\ge 0$.

The self-adjoint extensions of the operator $A_0$ are
parameterized by (\ref{exten}).   For $s=1$ the domain $D(A_0)$ is
composed of regular (for example, the regular Bessel
$J_{\nu}(\sqrt{\lambda}r)$) functions for the range of the
extension parameter $0<\theta<\pi/2$. Because of the invariance of
Eq. (\ref{e13}) under simultaneous transformations $\mu\to-\mu$,
$l\to -l$ and $s\to-s$ we shall restrict ourselves to study of the
case $\mu>0$.

Therefore, for $s=1$  the complete set of functions of the
self-adjoint Hamiltonian is given by the regular functions defined
on the real half-line \bb f(r, \varphi)=
J_{|l+\mu|}(\sqrt{\lambda} r)\exp(il\varphi),\quad l=0, \pm 1, \pm
2, \ldots .
 \label{regsol}\ee

For $s=-1$ for the range of the extension parameter
$\pi/2<\theta<3\pi/2$ we derive that the defect subspaces contain
two singular Bessel functions $J_{-\gamma}(\sqrt{\lambda}r)$ and
$J_{\gamma-1}(\sqrt{\lambda}r)$ defined on the real half-line but
bearing in mind that $0<\gamma<1$ we have to leave only one of
them. We shall leave $J_{-\gamma}(\sqrt{\lambda}r)$. In order to
the set of functions would be complete in $l$, we need to remove
from the domain $D(A_0)$ the function
$J_{\gamma}(\sqrt{\lambda}r)$ with $l=-[\mu]$. Hence, the complete
set of functions of the self-adjoint Hamiltonian is given by  \bb
f(r, \varphi)= J_{|l+\mu|}(\sqrt{\lambda} r)\exp(il\varphi),\quad
l=0, \pm 1, \pm 2, l\ne -[\mu], \ldots ;
J_{-\gamma}(\sqrt{\lambda} r)\exp(-i[\mu]\varphi).
 \label{regs-1}\ee

Therefore, the complete set of functions of a
self-adjoint Hamiltonian for the Pauli equation in the AB
potential is a set (\ref{regsol}) for $s=1$ and
a set (\ref{regs-1}) for $s=-1$, in which
we must replace in the argument of Bessel functions
$\sqrt{\lambda}$ by $k\equiv\sqrt{2mE}$.

In addition, if the self-adjoint extension of $A_0$ is selected by
 the range of the extension parameter
$\pi/2<\theta<3\pi/2$ there appears a bound state, which is
expressed through the MacDonald function $K_{(1-\gamma)}(z)$ (or
$K_{\gamma}(z)$). The appearance of bound state ``suggests that this
range of parameters in the effective Hamiltonian parameterizes
nontrivial physics in the core'' \cite{phg}. One sees from our model
the appearance of bound state  is possible only if  the additional
spin term in Eq. (\ref{e13}) is attractive, i.e. $s=-1$ (see, also
\cite{khal,khal1}).

\section{Scattering of  electrons by an AB
field}

We shall consider in what follows the case
\bb \mu=[\mu]+\gamma\equiv n+\gamma, \label{divi}\ee where $n$ is
an integer and \bb 1>\gamma>0. \label{nonint}\ee

The expansion of the spatial electron wave function is given by  \bb
 \psi(r, \varphi)=\sum\limits_{l=-\infty}^{\infty}
N_lJ_{\nu}(kr)e^{il\varphi},\quad s=1,  \label{1newsc0}
 \ee
where $N_l$ are constants and
\bb
 \psi(r, \varphi)=\sum\limits_{l=-\infty}^{\infty}
N_lJ_{\nu}(kr)e^{il\varphi} + D_{l=-n}J_{-\gamma}(kr)e^{-in\varphi} ,\quad
s=-1.
\label{1newsc}
 \ee
Here the summation is carried out with the omission of the $l=-n$
term.

If we assume that the electron wave propagates from the left along
the $x$ axis so as to incident wave is $\psi=e^{ikx}$, then
$\varphi$ is the scattering angle measured from the positive $x$
axis. The asymptotic form of electron wave function at $r\to
\infty$ is a superposition of ingoing plane wave and scattered
outgoing cylindrical wave
 \bb
 \psi_k(r, \varphi) = e^{ikx}
+\frac{f(\varphi)}{\sqrt{r}}e^{i(kr-\pi/4)}. \label{solscat}
 \ee
At $r\to \infty$ the plane wave is given by
\bb
 e^{ikr\cos\varphi}\to \frac{1}{\sqrt{2\pi kr}}\sum\limits_{l=-\infty}^{\infty}
e^{il\varphi}\left(e^{i(kr-\pi/4)} + (-1)^le^{-i(kr-\pi/4)}\right)
\label{exppw}
 \ee
and Eq. (\ref{1newsc0}) is given with using the expansion of Bessel functions as follows:
\bb
 \psi(r, \varphi)\to\frac{1}{\sqrt{2\pi kr}} \sum\limits_{l=-\infty}^{\infty}
N_l e^{il(\varphi+\pi/2)}\left(e^{i(kr-\pi/4)}e^{-i\pi|l+\mu|/2} +
e^{-i(kr-\pi/4)}e^{i\pi|l+\mu|/2}\right).
\label{ABsol}
 \ee
Choosing the coefficients $N_l$ so as to the ingoing waves would
be canceled in Eq. (\ref{solscat}), one easily obtains \bb
N_l=e^{-i(\pi/2)|l+\mu|}. \label{1ampscat0} \ee

The scattering amplitude for solution (\ref{1newsc0}) is then found to 
be given by the AB formula
 \bb
f_1(\varphi)= \sin\pi\gamma\frac{e^{-i(n+1/2)\varphi}} {\sqrt{2\pi
k}\sin(\varphi/2)}. \label{ABamp}
 \ee
Inserting in (\ref{1newsc})
the $(\pm)\exp(-i\pi(\mu-n)/2)J_{\gamma}(kr)$ terms we obtain
the scattering amplitude in the AB form again
\bb
 f_{-1}(\varphi)= \sin\pi\gamma\frac{e^{-i(n-1/2)\varphi}}
{\sqrt{2\pi k}\sin(\varphi/2)}.
 \label{1ampsct}
 \ee

Therefore, if electrons are polarized so as to
their spins are oriented along the $z$ axis the scattering
cross section in an AB potential is given by the AB formula
 \bb
\frac{d\sigma}{d\varphi} = \frac{\sin^2\pi \gamma} {2\pi k
\sin^2(\varphi/2)}. \label{ABsec}
 \ee
The cross section for unpolarized electrons is obviously described
Eq. (\ref{ABsec}) too.

It is easily to see that when the electron spin is perpendicular
to the $z$ axis we must take the wave function in the form \bb
 \psi(r, \varphi)=\sum\limits_{l=-\infty}^{\infty}
N_lJ_{\nu}(kr)e^{il\varphi} + e^{-in\varphi}(J_{\gamma}(kr)e^{-i\pi(\mu-n)/2}+J_{-\gamma}(kr)e^{i\pi(\mu-n)/2})/2,
\label{newss0}
 \ee
where the summation is carried out with the omission of the $l=-n$
term. 

Now the scattering amplitude has the form
\bb
 f_0(\varphi)= \frac{\sin\pi\gamma}{\sqrt{2\pi k}}\left(\frac{e^{-i(n+1/2)\varphi}}
{\sin(\varphi/2)} +ie^{-in\varphi}\right)
 \label{sctas0}
 \ee
and the cross section is
 \bb
\frac{d\sigma}{d\varphi} = \frac{\sin^2\pi \gamma} {2\pi k}
\left(\frac{1}{\sin^2(\varphi/2)} -1\right) \label{1secsct}
 \ee
in case the electron spin lies in the plane of scattering. In case
$|\gamma|<1$ this formula describes the scattering cross section
of spin-polarized electrons by a long cylindrical magnetic flux
tube of small radius \cite{khal}.

\subsection{On a bound electron state in the AB field}

It was shown in Section II that in the range of the extension parameter
$\pi/2<\theta<3\pi/2$ there appears a bound state, which is
expressed through the MacDonald function $K_{(1-\gamma)}(z)$ (or
$K_{\gamma}(z)$). The radial part of wave function of bound state is
solution of Eq. (\ref{e13}) with $s=-1$ and $E<0$. It is $K_{(1-\gamma)}(z)$ (or
$K_{\gamma}(z)$) of argument $z=r\sqrt{2m|E|}$ defined on the real
half-line where $m$ is the electron mass and $E<0$ is the energy
of bound state.
One sees that a bound electron state may occur
in the quantum system under consideration if the interaction of
the electron spin with a magnetic field is included and it is attractive.

\section{Scattering of  a massive neutral fermion in the AC
field configuration}

If we put $|e|=Ms$ in equation (\ref{e13}) it will describe the
motion of a massive neutral fermion with the anomalous magnetic
moment $M$ in the electric field of a linear charge aligned
perpendicularly to the fermion motion. Then, $\nu$ is
 \bb
 \nu =|l+s\gamma|\ne 0,
\quad \gamma=MB, \quad k = \sqrt{2m E}, \label{fnotion12}
 \ee
where $B/2$ is the linear charge density, $s=\pm 1$ is the fermion spin projection on the
$z$ axis, $E$ is the fermion energy and  the results obtained above are
valid for a neutral fermion with AMM in the AC configuration.

 The scattering amplitudes  for $\mu>0,\quad \mu<0,\quad  s=\pm 1$ are respectively  \cite{khm} \bb
 f_s(\varphi)=
\frac{s}{\sqrt{2\pi k}}\frac{e^{is(n+1/2)\varphi+in\pi}
\sin(\pi\gamma)}{\sin(\varphi/2)}, \label{fampscat}
 \ee

 \bb
 f_s(\varphi)=
-\frac{s}{\sqrt{2\pi k}}\frac{e^{is(|n|-1/2)\varphi+i|n|\pi}
\sin(\pi|\gamma|)}{\sin(\varphi/2)}. \label{facscat2}
 \ee
Thus, if the spin of neutral fermion is oriented along the $z$
axis the scattering cross section in an the AC configuration is
given by \bb \frac{d\sigma}{d\varphi}_{AC} = \frac{\sin^2(\pi
\gamma)} {2\pi k \sin^2\varphi/2}. \label{ACsec}
 \ee
The cross section for unpolarized neutral fermions with AMM in the
AC configuration is also described by Eq. (\ref{ACsec})
\cite{ahc}. Eqs. (\ref{fampscat}), (\ref{facscat2}) and
(\ref{ACsec}) coincide respectively with Eqs. (\ref{ABamp}),
(\ref{1ampsct}) and (\ref{ABsec}).

If the spin of neutral fermion in the initial state lies in the
scattering plane then the scattering amplitude and cross section
are respectively
 \bb
 f_0(\varphi)=
(-1)^{n+1}\frac{\sin(\pi\gamma)}{\sqrt{2\pi
k}\sin(\varphi/2)}\sin(n+1/2)\varphi, \label{ACamsc}
 \ee
and
 \bb
 \frac{d\sigma}{d\varphi} =|f_0(\varphi)|^2=
\frac{\sin^2(\pi \gamma)} {2\pi
k\sin^2(\varphi/2)}\sin^2(n+1/2)\varphi~.\label{ACsecsct}
 \ee
One can show that these formulas are described the case of
negative $\mu$ too, if  $n$ replaced by $|n|-1+\theta(\mu)$, where
$\theta(x)=[1+{\rm sign}(x)]/2$. It is seen if  the particle spin
in the initial state lies in the scattering plane the scattering
amplitudes and cross sections for AB and AC effects are different.

\section{Discussion}

If the dependence of cross section is to be studied upon the spin
polarization of particles in the initial state, the detector
should detect only those scattered particles for which the angles
between the spin and momentum vectors are equal in initial and
final states. In the AB configuration the magnetic field ${\bf H}$
is localized inside infinitely thin solenoid, so the Hamiltonian
(\ref{e12}) does not contain the vector $\bm{\sigma}$ matrix in
the range $r>0$ and the electron spin ${\bf s}$ oriented along a
unit vector ${\bf n}$ is conserved. Since, the scattering occurs
in the $xy$ plane the spin vector of electron in initial state can
effect the scattering amplitude and cross section only when it has
a nonzero projection on the plane of scattering. So, in fact, the
dependence on the (initial) spin of electron in the cross section
determines and arises because the direction of motion of electron
is changed as a result of scattering (see \cite{khal1}).

In case a fermion with AMM moves in the $xy$ plane, the
Hamiltonian contains only the $\sigma_3$ matrix. Therefore, only
the fermion spin projection on the $z$ axis $s$ is conserved. Like
the AB case the spin vector of fermion in initial state can effect
the scattering amplitude when it has a nonzero projection on the
scattering plane but in the AC case the dependence on the spin
vector in the initial state in the cross section determines not
only by the change in momentum of fermion but also by the change
in the spin orientation of scattered fermion (the quasiclassical
precession of fermion spin).

\vskip 1truecm

\begin{acknowledgments}

 This paper was supported  in part by the Program for Leading Russian
Scientific Schools (Grant No. NSh-3312.2008.2).

\end{acknowledgments}


\begin{thebibliography}{55}


\bibitem{1} Y. Aharonov and D. Bohm, Phys. Rev., {\bf 115} (1959) 485.

\bibitem{aw} M.G. Alford and F. Wilczek, Phys. Rev. Lett., {\bf 62} (1989) 1071.

\bibitem{jmpt} R. Jackiw, A.I. Milstein, S.-Y. Pi, and I.S.
Terekhov, {\sl Induced Current and Aharonov--Bohm Effect in
Graphene}, e-print arXiv:cond-mat.mes-hall/0904.2046v3.

\bibitem{ivt} I.V. Tyutin, {\sl Electron Scattering by a Solenoid},
Preprint of P.N. Lebedev Institute, No 27 (1974), unpublished;
e-print arXiv:quant-ph/0801.2167 v2.

\bibitem{phg} Ph. Gerbert, Phys. Rev.,  {\bf D40} (1989) 1346.

\bibitem{ampw} M.G. Alford, J. March-Pussel and F. Wilczek,
Nucl.Phys., {\bf B328} (1989) 140.

\bibitem{phgj} Ph. Gerbert and R. Jackiw, Commun. Math. Phys.,  {\bf 124} (1989) 229.

\bibitem{khnew} V.R. Khalilov, Theoretical and Mathematical Physics (TMP), in press.

\bibitem{cfp} F.A.B. Coutinho and J. Fernando Perez, Phys. Rev.,
{\bf D49} (1994) 2092.

\bibitem{ggs} S.P. Gavrilov, D.M. Gitman, and A.A. Smirnov,
{\sl Dirac equation in a magnetic-solenoid field}, e-print arXiv:
hep-th/0210312v3.

\bibitem{ggsv} S.P. Gavrilov, D.M. Gitman, A.A. Smirnov, and B.L.
Voronov, {\sl Dirac fermions in a magnetic-solenoid field},
e-print arXiv:hep-th/0308093v2.

\bibitem{ahc} Y. Aharonov and A. Casher, Phys. Rev. Lett., {\bf 53} (1984)
319.

\bibitem{erbm} E.R. Bezerra de Mello,
{\sl Effects of Anomalous Magnetic Moment in the Quantum Motion of
Neutral Particle in Magnetic and Electric Fields Produced by a
Linear Source in a Conical Spacetime}, e-print arXiv:
hep-th/0403022v3.


\bibitem{khal} V.R. Khalilov, C.L. Ho, Ann. Phys., {\bf 323} (2008) 1280.


\bibitem{nai} M.A. Naimark, {\sl Theory  of Linear Differential operators}, (Nauka, Moscow, 1969).

\bibitem{crein} {\sl Functional Analysis}, $2^{nd}$ ed., Ed: S.G. Krein (Nauka, Moscow, 1972).

\bibitem{GR} I.S. Gradshteyn and I.M. Ryzhik, {\sl Table of Integrals, Series,
and Products}, $5^{th}$ ed., (Academic Press, San Diego, 1994).

\bibitem{olv} F.W.J. Olver, {\sl Introduction to Asymptotics and
Special Functions}, (Academic Press, New York and London, 1974).

\bibitem{khal1} V.R. Khalilov, Mod. Phys. Lett., {\bf A21} (2006) 1647.


\bibitem{khm} V.R. Khalilov, I.V. Mamsurov, TMP, {\bf 161} (2009) 1503.




\end{thebibliography}
\end{document}